\documentclass[slac_one]{revtex4}
\usepackage{graphicx}
\usepackage{fancyhdr}
\usepackage{amsmath}
\renewcommand{\Re}{{\rm Re\thinspace}}
\renewcommand{\Im}{{\rm Im\thinspace}}

\newcommand{\Lumint}{{\cal L}_{\rm int}}

\def\PE{{\bf P}_{e^-}}
\def\PP{{\bf P}_{e^+}}

\def\PET{{\bf P}_{e^-}^T}
\def\PPT{{\bf P}_{e^+}^T}

\begin{document}

\title{{\small{2005 International Linear Collider Workshop - Stanford,
U.S.A.}}\\ 
\hfill {\rm\small CERN-PH-TH/2005-127\\[-.5ex] \ \hfill
\\}%
\vspace{6pt}
Positron Polarization at the International Linear Collider}

\author{P. Osland}
\affiliation{CERN, CH-1211 Geneva 23, Switzerland  and University of Bergen, 
N-5007 Bergen, Norway}
\author{N. Paver}
\affiliation{University of Trieste and INFN, I-34100 Trieste, Italy}

\begin{abstract}
We review some recent arguments supporting the upgrade of the 
International Linear Collider by a polarized positron beam, in addition to 
the polarized electron beam. The examples presented here mainly focus on the 
impact of positron polarization on items relevant to new physics searches, 
such as the identification of novel interactions in fermion-pair 
production and the formulation of new CP-sensitive observables. In 
particular, in addition to the benefits from positron and electron 
longitudinal polarizations, the advantages in this field of having 
transverse polarization of both beams are emphasized.   
\end{abstract}

\maketitle

\thispagestyle{fancy}

\section{INTRODUCTION} 
Electron beam longitudinal polarization at the SLC has been a really powerful
tool to test the structure of the Standard Model (SM) electroweak interactions
and to make precise measurements of the relevant parameters, in particular of
the electroweak mixing angle through the left-right asymmetry.  No doubt the
electron beam with high degree of polarization, already foreseen, will play a
similar important r\^ole at the International Linear Collider (ILC). In
addition to scrutinizing the SM dynamics, it should enhance the experimental
sensitivity to new physics (NP), which is considered as one of the major parts 
of the physics programme at this machine. The new physics (NP) might manifest
itself either directly {\it via} the production of new particles whose
properties and quantum numbers must be tested, or indirectly through small
deviations of cross sections from the SM predictions caused by novel
interactions mediated by very heavy states whose couplings must be determined
(or constrained) in precision measurements.

In this regard, the full potential of the International Linear Collider  
could be exploited only with the additional facility of a polarized positron 
beam. Firstly, at a given annual luminosity, polarized cross sections would 
be enhanced, in particular an effective polarization larger that the 
individual $e^-$ and $e^+$ polarizations can be defined and more accurately 
determined. This would greatly increase the sensitivity to small, 
non-standard, cross sections or to deviations from the SM cross sections. 
Secondly, the efficiency for chasing non-standard particles 
(such as, for example, the SUSY particles or the Kaluza-Klein gravitons) 
could be dramatically improved by strong SM background suppression 
accompanied by signal enhancement, as allowed by appropriately tuning 
electron and positron polarizations. Finally, more transition amplitudes and 
corresponding polarized observables could be defined, which would allow to 
disentangle the independent components of the novel interactions'
multi-parameter space and explore it in detail {\it via} largely 
model-independent approaches. 

\begin{figure*}[t]
\centering
\includegraphics[width=120mm]{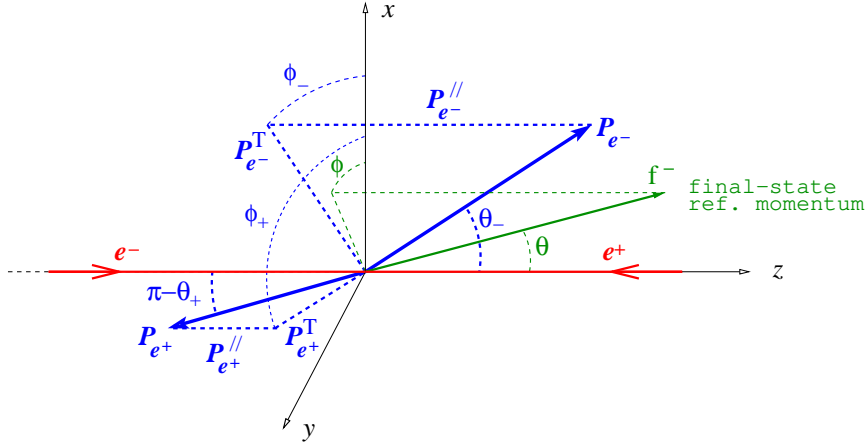}
\caption{Coordinate system used for momenta and polarization vectors.}
\label{fig:angles}
\end{figure*}

A strong case can also be made for the opportunity of having both beams
transversely polarized, which should be realized by the application of spin
rotators to the longitudinally polarized beams. By providing the possibility
of measuring new (azimuthal) angular distributions and asymmetries, this would
extend substantially the potential of searches for novel sources of CP
violation and of anomalous triple gauge couplings, as well as to discriminate
among models of gravity in extra dimensions. The coordinate system, as well as
the angles defining transverse polarization (longitudinal polarization is
directed along the $e^+-e^-$ axis), are shown in Fig.~\ref{fig:angles}.  The
examples exposed in the sequel are taken from the exhaustive report on the
physics potential of the ILC with both electron and positron beams polarized
\cite{Moortgat-Pick:2005cw}.

\section{TOP-QUARK FLAVOUR CHANGING NEUTRAL COUPLINGS}
FCN {\it top} couplings are predicted in numerous SM extensions, and therefore
represent an interesting field in NP searches, also because they can only
occur in the SM through strongly GIM suppressed loops. The $t\to Vq$
transitions, with $V=\gamma,Z$ and $q=u,c$, are generally described by an
effective interaction, comprising both $\gamma_\mu$- and
$\sigma_{\mu\nu}$-couplings, see for example
\cite{Aguilar-Saavedra:2001ab}. Reactions sensitive to FCN couplings are
either $t\bar t$ production $e^+e^-\to t{\bar t}\to VqW^-{\bar b}$, or single
top production $e^+e^-\to{\bar t}q\to W^-{\bar b}q$. Examples of the
three-$\sigma$ reach on FCN {\it top} branching ratios obtainable with
different longitudinal beam polarization and with c.m.\ energy and
time-integrated luminosity $E_{\rm c.m.}=0.5$ TeV, ${\cal L}_{\rm int}=300$
${\rm fb}^{-1}$ and $E_{\rm c.m.}=0.8$ TeV, ${\cal L}_{\rm int}=500$ ${\rm
fb}^{-1}$, are reported in Tab.~I
\cite{Aguilar-Saavedra:2001ab,Glover:2004cy}. The improvement from $e^+$
polarization is significant. Comparing with hadron colliders, the LHC (with
limits foreseen in the $10^{-5}$ range) and the ILC may be considered as
complementary, with the ILC superior in the discovery reach on
$\sigma_{\mu\nu}$-couplings \cite{Glover:2004cy}.
\begin{table}[htb]
\begin{center}
\caption{3$\sigma$ limits on FCN top branching ratios} 
\begin{tabular}{|l|c||c|c|c|c|}
\hline
&\raisebox{-1.5ex}{[$10^{-4}$]} & $E=500$~GeV &  $E=500$~GeV  
& $E=500$~GeV &  $E=800$~GeV \\
& & unpol & (80\%,0) & (80\%,45\%) & (80\%,60\%) \\
\hline \hline
$\gamma_\mu$ & ${\rm BR}(t\to Zq)$& 6.1 & 3.9 & 2.2 & 1.9\\ \hline
$\sigma_{\mu\nu}$ & ${\rm BR}(t\to Zq)$& 0.48 & 0.31 & 0.17 & 0.07\\ \hline
$\gamma_\mu$ & ${\rm BR}(t\to \gamma q)$& 0.30 & 0.17 & 0.093 & 0.038\\ \hline
\end{tabular}
\end{center}
\end{table} 
\section{CP-VIOLATON IN 3- AND 4-JET Z-BOSON DECAYS}
This sector is promising for searches of new CP-violation sources at the GigaZ
option, due to the SM being extremely suppressed there, and particularly
appealing are the transitions $Z\to{\bar b}bg$ and $Z\to{\bar b}bgg$, leading
to 3-jet and 4-jet final states ($g$ denotes the gluon) owing to the
excellent $b$ (and $\bar b$) tagging expected at the ILC.  A model-independent
parameterization of such processes can be obtained by the effective Lagrangian
\begin{equation}
{\cal L}_{\rm CP}=\left[h_{Vb}\hskip 3pt {\bar{b}}T^a\gamma^{\nu}b +  
h_{Ab}\hskip 3pt{\bar{b}}T^a\gamma^{\nu}\gamma_5\right]\hskip 2pt 
Z^{\mu}G^a_{\mu\nu}, 
\nonumber
\end{equation}
with $T^a$ the color matrices and $G_{\mu\nu}^a$ the familiar gluon
field-strength tensor.  Dimensionless constants ${\hat h}_{Vb}$ and ${\hat
h}_{Ab}$ can be defined out of $h_{Vb}$ and $h_{Ab}$ by essentially using
$m_Z^2$ as a scale parameter. Different new physics models envisage the
coupling constants $h_{Vb}$ and $h_{Ab}$, in particular the quark-substructure
scenario
\begin{equation}
{\cal L}^\prime = \frac{e}{2\sin\theta_W\cos \theta_W}\hskip 2pt 
Z_{\mu}{\bar{b}}^\prime\gamma^{\mu}\left(g_V^\prime-g_A^\prime \gamma_5\right) 
b - i\frac{g_s}{2 m_{b^\prime}}\hskip 2pt \hat{d}_c 
{\bar{b}}^\prime\sigma^{\mu\nu}\gamma_5T^a b G^a_{\mu\nu} + {\rm h.c.} 
\nonumber
\end{equation} 
Here, $g_V^\prime$, $g_A^\prime$ and ${\hat d}_c$ are (in principle complex)
constants expected of order unity according to compositeness arguments,
$b^\prime$ denotes a $b$-quark excitation and it is understood that
$m_{b^\prime}\gg m_Z$. The constants ${\hat h}_{Vb}$ and ${\hat h}_{Ab}$ are
determined by virtual $b^\prime$ exchange. One can define different CP
asymmetries, such as $T_{33} = (\widehat{\bf k}_{\bar{b}} - \widehat{\bf
k}_b)_3 (\widehat{\bf k}_{\bar{b}} \times \widehat{\bf k}_b)_3$ and $V_3 =
(\widehat{\bf k}_{\bar{b}} \times \widehat{\bf k}_b)_3$, where `3' corresponds
to the $e^+$ direction \cite{Nachtmann:2003hg}. We show in Tab.~II the
one-$\sigma$ accuracy obtainable at the ILC from $V_3$ on the relevant
dimensionless constant ${\hat h}_b$ from $Z\to 3$-jet events
\cite{Nachtmann:2003hg}.
\begin{table}[htb]
\begin{center}
\caption{Accuracy on ${\hat h}_b$ and lower limits on $m_{b^\prime}$} 
\begin{tabular}{|c|c|c|c|c|c|c|}
\hline
\multicolumn{2}{|c}{}  & \multicolumn{5}{|c|}{$(P_{e^-},P_{e^+})$} \\ 
$V_3$ & $y_{\rm cut}$ & $(0,0)$ & $(+80\%,0)$ & $(-80\%,0)$ &
$(+80\%,-60\%)$ & $(-80\%,+60\%)$ \\\hline
$\widetilde{h}_b$ $[10^{-3}]$
& 0.01 & 8.5 & 2.5 & 1.9 & 1.8 & 1.4 \\ 
& 0.1  & 9.5 & 3.0 & 2.1 & 2.0 & 1.5 \\ \hline
$m_{b'}$ [TeV]  & 0.01 & 0.9 & 1.6 & 1.9 & 2.0 & 2.2 \\
                  & 0.1  & 0.9 & 1.5 & 1.8 & 1.9 & 2.1 \\ \hline 
\end{tabular}
\end{center}
\end{table}
With both beams polarized, the existing sensitivity from LEP is improved by 
about an order of magnitude, and also the bounds on $m_{b^\prime}$ are 
significant, current limits on quark excitations being quite poor.  




\section{MODEL-INDEPENDENT CONTACT INTERACTION SEARCHES}
Contact interactions (CI) are used as ``low energy'' representations 
of NP dynamics characterized by exchanges of objects with mass scales 
$\Lambda$ (much) heavier than the available c.m. energy, that can accordingly 
be signalled only indirectly, {\it via} deviations of cross sections from the 
SM predictions. For $e^+e^-\to{\bar f}f$, we consider the effective, 
dimension-6 contact interaction Lagrangian \cite{Eichten:1983hw}
\begin{equation}
{\cal L}_{\rm CI}
=\frac{1}{1+\delta_{ef}}
\sum_{i,j={\rm L,R}}\frac{4\pi\eta_{ij}}{\Lambda^2_{ij}}
\left(\bar e_{i}\gamma_\mu e_{i}\right)
\left(\bar f_{j}\gamma^\mu f_{j}\right), 
\label{lagra}
\nonumber
\end{equation} 
with $\vert\eta_{ij}\vert=1,0$. Although originally inspired by fermion
compositeness remnant binding forces, this current-current CI may
well-represent new physics induced by exchanges of objects heavier than $\sqrt
s$ and $\sqrt{|t|}$ of the process under consideration, for example
$Z^\prime$s and leptoquarks. In general, $\sigma$ and $A_{\rm FB}$ depend on
all CI couplings, unless one considers specific models where only one of them
is assumed non-zero. Longitudinal polarization of both beams allows the
definition of more observables, and is therefore decisive to perform a
model-independent data analysis where the different couplings can be
considered as simultaneously non-zero independent free parameters and yet be
separately tested or constrained. The resulting 95\% C.L. separate reaches on
the mass scales $\Lambda$, obtainable from the ${\bar b}b$ and ${\bar c}c$
channels at the ILC with $E_{\rm c.m.}=0.5$ TeV as a function of the
luminosity, are shown in Fig.~2 \cite{Babich:2001nc}. Thin lines represent
there ($\vert P_{e^-}\vert, \vert P_{e^+}\vert$)=(80\%,0), while thick lines
represent ($\vert P_{e^-}\vert, \vert P_{e^+}\vert$)=(80\%,60\%). The great
impact of positron polarization is evident in this figure (for related
model-dependent analyses see \cite{Riemann:2001bb}).
\begin{figure*}[htb]
\centering 
\includegraphics[width=75mm]{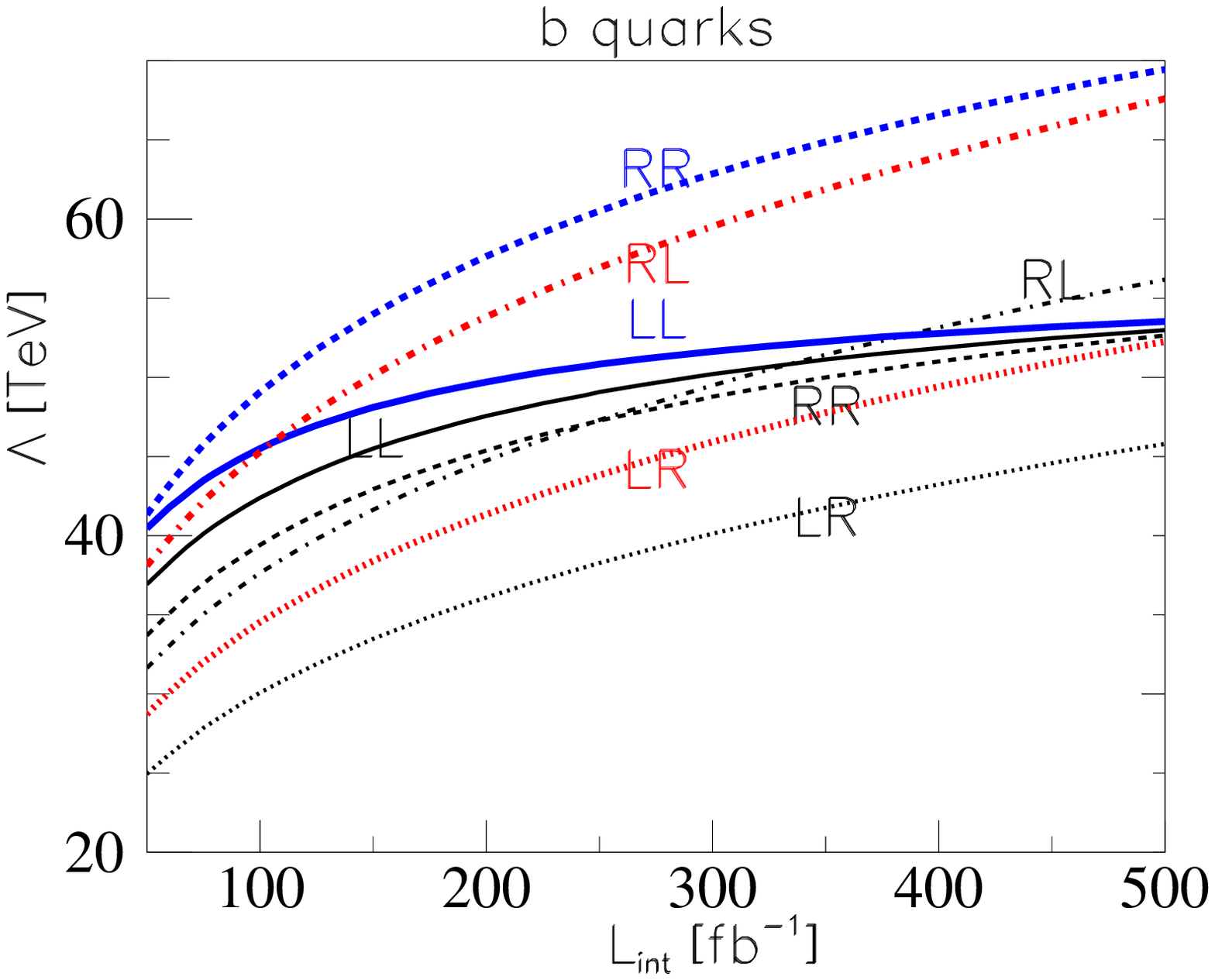}
\includegraphics[width=75mm]{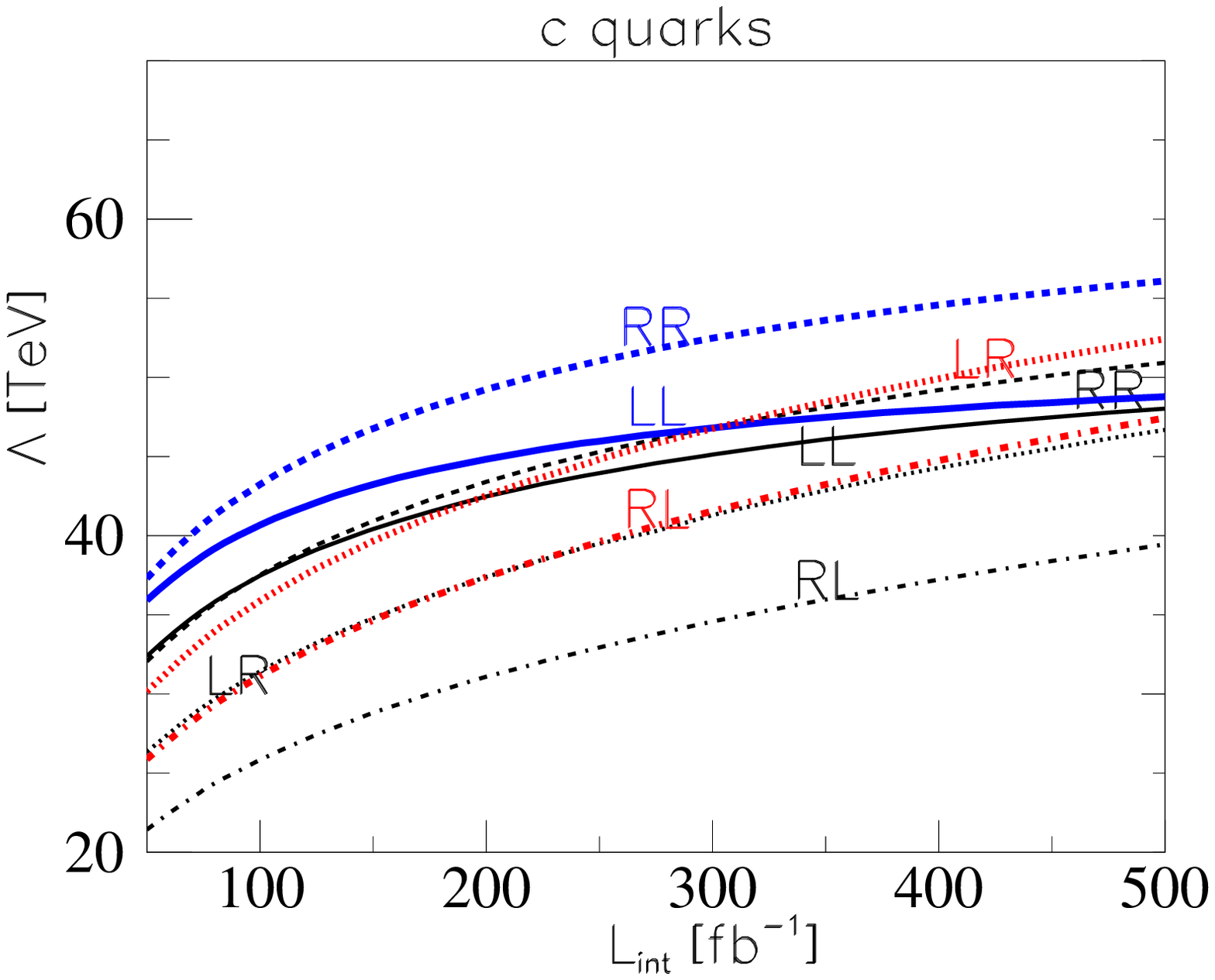}
\caption{95\% C.L.\ lower bounds on the CI mass scales $\Lambda$.}
\end{figure*} 

\section{TRANSVERSELY POLARIZED BEAMS}
For the reaction $e^+e^-\to{\bar f}f$, with $\bf p$ and $\bf k$ the momenta of 
the $e^-$ and $f$ momenta, one can write the transversely polarized 
differential cross section as 
\begin{align}
d\sigma_\text{pol}
&=A_1[\PPT\cdot{\bf k}+\PET\cdot{\bf k}]
+A_2[\PPT\cdot{\bf k}-\PET\cdot{\bf k}]+B(\PPT\cdot\PET)
+C(\PPT\cdot{\bf k})(\PET\cdot{\bf k})
+D(\PPT\times\PET)\cdot{\bf p} \nonumber \\
&+E_1[(\PPT\times{\bf k})\cdot {\bf p}](\PET\cdot{\bf k})
 +E_2[(\PET\times{\bf k})\cdot {\bf p}](\PPT\cdot{\bf k})
\nonumber
\end{align}
It can be easily seen that the terms $A,D$ and $E$ violate P and, with
reference to Fig.~\ref{fig:angles}, the terms $B$ and $C$ combine to give a
dependence $\propto |\PET | |\PPT |\sin^2\theta\cos
\left(2\phi-\phi_--\phi_+\right)$, while the $E$ terms give the dependence 
$\propto |\PET | | \PPT | \sin^2\theta\sin
\left(2\phi-\phi_- -\phi_+\right)$.
Clearly, the advantage of transverse polarization is that either $\PPT$ or
$\PET$ provides a new direction, and this allows measuring CP violation without
(sometimes complicated) final-state spin analyses. Also, with only $(V,A)$
couplings, the $A$ terms linear in the transverse polarizations vanish for 
$m_e\to 0$ \cite{Dass:1975mj}.
\subsection{New sources of CP-violation}
Unconventional interactions in $e^+e^-\to {\bar t}t$ can be mediated by
(pseudo)-scalar or tensor effective {\it top} quark couplings. Such new,
helicity changing, currents can at the leading order interfere with the SM
$s$-channel ($V,A$) $\gamma$ and $Z$ exchanges also as $m_e\to 0$, and allow
the definition of CP-odd observables that can be measured if transverse beam
polarization is available \cite{Ananthanarayan:2003wi}. One can introduce an
effective Lagrangian representation of these new interactions in terms of a
(high) mass scale $\Lambda$ and dimension-6 four-fermion operators ${\cal
O}_k$ times coefficients $\alpha_k$ of order unity. The relevant piece, to be 
added to the SM interaction, can be represented as:
\begin{equation}
{\cal L}^{4F}=\frac{1}{\Lambda^2}
\sum_k\left(\alpha_k{\cal O}_k+{\rm h.c.}\right)
 =\sum_{i,j={\rm L,R}}\left[S_{ij}\left(\bar{e}P_ie\right)
\left(\bar{t}P_jt\right) 
+V_{ij}\left(\bar{e}\gamma_{\mu}P_ie\right)
\left(\bar{t}\gamma^{\mu}P_jt\right) 
+T_{ij}\left(\bar{e}\frac{\sigma_{\mu\nu}}{\sqrt{2}}P_ie\right)
\left(\bar{t}\frac{\sigma^{\mu\nu}}{\sqrt{2}}P_jt\right)\right]
+{\rm h.c.}
\nonumber
\end{equation}
Here, $P_i=\frac{1}{2}\left(1\pm\gamma_5\right)$ and $S_{ij}$, {\it etc.} are 
a priori complex constants. Up-down CP-odd asymmetries in the azimuthal angle 
$\phi$ can be constructed, for example the one essentially sensitive to 
Im$S_{\rm RR}$ \cite{Ananthanarayan:2003wi}:
\begin{equation}
A(\theta_0)=\frac{1}{\sigma^{+-}}
\int_{-\cos\theta_0}^{\cos\theta_0}d\cos\theta
\left[\int_0^\pi \frac{ d\sigma^{+-}}{d\Omega} d\phi 
- \int_{\pi}^{2\pi} \frac{d\sigma^{+-}}{d\Omega} d\phi
\right],
\nonumber
\end{equation}
where the superscripts denote aligned opposite transverse polarizations of 
$e^-$ and $e^+$, and $\theta_0$ defines a cut-off around the beam directions. 
The 90\% C.L. sensitivity on Im$S$ for $E_{\rm c.m.}=0.5$ TeV, 
${\cal L}_{\rm int}=500$ ${\rm fb}^{-1}$ and full (i.e., 100\%) transverse 
polarization, can be as low as $10^{-3}$, which translates to the mass scale 
$\Lambda\sim 8$ TeV.  

Another representative example of the benefits of positron and electron 
transverse polarization is the search for anomalous, CP-violating 
$\gamma\gamma Z$ and $\gamma ZZ$ couplings in the production process 
$e^+e^-\to\gamma Z$ at the ILC. In this case, ($V,A$) couplings of the 
(self-conjugate) $\gamma$ and $Z$ occur both in the $s$- and in the 
$t$-channel, and with transverse polarization CP-odd interference of 
anomalous couplings with the SM ones is allowed to occur. One can define the 
anomalous vertex in terms of, a priori complex, couplings $\lambda_1$ and 
$\lambda_2$ as follows \cite{Abraham:1993zh}: 
 \begin{equation}
{\cal L} = 
   e \frac{{\lambda_1}}{ 2 m_Z^2} F_{\mu\nu}
    \left( \partial^\mu Z^\lambda \partial_\lambda Z^\nu
          - \partial^\nu Z^\lambda \partial_\lambda Z^\mu
      \right)
      +\frac{e}{16 c_W s_W} \frac{{\lambda_2}}{m_Z^2}
       F_{\mu\nu}F^{\nu \lambda}
       \left(\partial^\mu Z_\lambda + \partial_\lambda Z^\mu   \right). 
\nonumber
\end{equation}  
Interference terms in the photon angular distribution depend on Re$\lambda_1$,
Re$\lambda_2$ and Im$\lambda_2$, and are proportional to the product of
electron and positron transverse polarizations (parallel polarization
directions can be assumed). These couplings can be tested by appropriately
defined CP-odd asymmetries, that exploit the dependences $\sim\cos\theta\cos
2\phi$ and $\sim\cos\theta\sin 2\phi$ and combine polar forward-backward with
azimuthal asymmetries \cite{Ananthanarayan:2004eb}. Explicit expressions can
be found in the original references, but it is useful to just mention that,
with $g_V$ and $g_A$ the familiar SM electron couplings: $A_1(\theta_0)\propto
-\PET\PPT g_A\hskip 3pt \Re \lambda_2$; $A_2(\theta_0)\propto\PET\PPT
\left[\left(g_V^2-g_A^2\right) \Im \lambda_1 - g_V \Im\lambda_2\right]$;
finally, $A_3(\theta_0) \propto \frac{\pi}{2}
\left[\left((g_V^2+g_A^2\right))\Im \lambda_1-g_V \Im\lambda_2 \right]+
\PET\PPT\left[\left(g_V^2-g_A^2\right)\Im \lambda_1-g_V
\Im\lambda_2\right]$. Consequently, taking into account these dependencies,
the couplings $\lambda_1$ and $\lambda_2$ can in principle be disentangled and
tested separately. Notice that the product of electron and positron transverse
polarizations enter in the asymmetries. Tab. III shows the 90\% C.L. limits
obtainable for an optimal cut-off $\theta_0=26^\circ$, $E_{\rm c.m.}=0.5$ TeV,
${\cal L}_{\rm int}=500$ ${\rm fb}^{-1}$ and transverse polarizations
$P^T_{e^-}$ and $P^T_{e^+}$ equal to 80\% and 60\%, respectively.
\begin{table}
\centering
\caption{90\% C.L. limits on anomalous couplings $\lambda_1$ and $\lambda_2$}
\begin{tabular}{l|c|c|c|c}
\hline
Coupling & \multicolumn{3}{|c|} {Individual limit from}
 & Simultaneous limits \\
& $A_1$ & $A_2$ & $A_3$ &\\ 
\hline 
Re $\lambda_2$ & $1.4\times 10^{-2}$ & & & \\
Im $\lambda_1$ && $6.2\times 10^{-3}$ &$ 3.8\times 10^{-3}$ & $7.1\times
10^{-3}$ \\
Im $\lambda_2$ && $9.1\times 10^{-2}$ &$ 3.0\times 10^{-2}$ & $6.7\times
10^{-2}$ \\
\hline
\end{tabular}
\end{table}
\subsection{Extra-dimensional gravity in fermion-pair production}
Indirect signals of TeV-scale gravity propagating in large, compactified,
extra dimensions can be searched for in $e^+e^-\to {\bar f}f$. In the ADD
scenario of gravity in extra spatial dimensions \cite{Arkani-Hamed:1998nn},
the exchange of a tower of Kaluza-Klein (KK) gravitons with almost continuous
equally spaced mass spectrum occurs, and can be represented by the dimension-8
effective interaction \cite{Hewett:1998sn}
\begin{equation}
{\cal L}^{\rm ADD}=i\frac{4\lambda}{M_H^4}T^{\mu\nu}T_{\mu\nu}, 
\nonumber
\end{equation}
where $\lambda=\pm 1$ and $M_H$ is a cut-off on the summation over the KK
states, expected in the TeV range. The $\cos\theta$-dependent deviations from
the SM, reflecting spin-2 exchange, are proportional to the graviton coupling
strength $f_G=\lambda\, E_{\rm c.m.}^4/(4\pi\alpha_{\rm e.m.}M_H^4)$.  In the
simplest version of the RS scenario \cite{Randall:1999ee}, space-time is
five-dimensional and the (narrow) spin-2 KK resonances can also be in the TeV
range but are unequally spaced. Formally, this can be accomplished by the
replacement, with $\Lambda_\pi$ of the TeV order:
\begin{equation}
\frac{\lambda}{M_H^4}\rightarrow\frac{-1}{8\Lambda_\pi^2}
\sum_{n}\frac{1}{s-m_n^2+im_n\Gamma_n}.
\nonumber
\end{equation}
Below the production threshold, indirect signals of graviton exchange can be
tested, and distinguished e.g. from the four-fermion contact interactions, by
using suitable polar asymmetries \cite{Osland:2003fn} or differential
distributions convoluted with Legendre polynomials
\cite{Rizzo:2002pc}. Tab.~IV-left shows, as an example, the five-$\sigma$
identification reach on $M_H$ {\it vs.} luminosity, obtainable at the ILC with
$E_{\rm c.m.}=0.5$ TeV with longitudinally polarized beams, and combining the
$f=\mu,\tau,c,b$ channels \cite{Osland:2003fn}. Clearly, the effect of the
additional positron polarization, although being mitigated by the high
dimension of the relevant effective operator, helps in increasing the
sensitivity to that parameter.
\begin{table}[htb]
\caption{Left: Identification reach on $M_H$. 
Center: $M_H$ reach. Right: ADD {\it vs.} RS distinction.}
\begin{center}
\begin{tabular}{|c||c|c|c|}
\hline
$M_H$ [TeV] & \multicolumn{3}{|c|}{$\Lumint [\text{fb}^{-1}]$}\\
$\sqrt{s}=0.5$~TeV (long. pol.)& 100 & 300 & 500 \\ \hline\hline
unpolarised beams & 2.3 & 2.6 & 2.9 \\ \hline
$(\PE,\PP)=(+80\%,0)$ & 2.5 & 2.8 & 3.05\\ \hline
$(\PE,\PP)=(+80\%,-60\%)$ & 2.45 & 3.0 & 3.25 \\ \hline
\end{tabular}
\hspace*{5mm}
\begin{tabular}{|c||c|c|c|c|}
\hline
$M_H$ [TeV] & \multicolumn{4}{|c|}{$\Lumint [\text{fb}^{-1}]$}\\
(transv. pol.)& 100 & 300 & 500 & 1000\\ \hline\hline
$\sqrt{s}=0.5$~TeV & 1.6 & 1.9 & 2.0 & 2.2 \\ \hline
$\sqrt{s}=0.8$~TeV & 2.4 & 2.6 & 2.8 & 3.1 \\ \hline
$\sqrt{s}=1.0$~TeV & 2.8 & 3.2 & 3.4 & 3.8 \\ \hline
\end{tabular}
\hspace*{5mm}
\begin{tabular}{|c||c|c|c|c|}
\hline
5 $\sigma$ disc. reach& \multicolumn{4}{|c|}{$\Lumint [\text{fb}^{-1}]$}\\
$M_H$ [TeV] & 100 & 300 & 500 & 1000\\ \hline\hline
$\sqrt{s}=0.5$~TeV & 1.2 & 1.3 & 1.4 & 1.6 \\ \hline
$\sqrt{s}=0.8$~TeV & 1.8 & 2.0 & 2.2 & 2.4 \\ \hline
$\sqrt{s}=1.0$~TeV & 2.2 & 2.4 & 2.6 & 2.8 \\ \hline
\end{tabular}
\end{center}
\end{table}
\par 
With transverse beam polarization, forward-backward azimuthal asymmetries 
can be defined \cite{Rizzo:2002ww}, but the identification power of such 
observables on $M_H$ is similar or less than obtained from longitudinally 
polarized beams. Tab.~IV-center shows the five-$\sigma$ reach from the 
combination of the $f=\mu,\tau,c,b$ and $t$ channels with transverse $e^-$ and 
$e^+$ polarizations of 80\% and 60\%, respectively. 

Conversely, with transverse polarization, discrimination among ADD and RS 
gravity scenarios is possible, by means of the azimuthal asymmetry between 
events at positive and negative values of $\sin 2\phi$ \cite{Rizzo:2002ww}: 
\begin{equation}
\frac{1}{N}\frac{d A_i^T}{d\cos\theta}=\frac{1}{\sigma}
\left[\int_+\frac{d\sigma}{d\cos\theta d\phi}-
\int_-\frac{d\sigma}{d\cos\theta d\phi}\right].
\nonumber
\end{equation} 
While vanishing for both the SM and the RS resonance cases (neglecting widths
with respect to masses), this asymmetry receives a finite contribution in the
ADD scenario, through an imaginary part that can be acquired by the graviton
coupling constant $f_G$. Tab.~IV-right shows the corresponding ADD {\it vs.} RS
five-$\sigma$ discrimination power at the ILC, for the same input values as in
Tab.~IV-center.

\begin{acknowledgments}
Research supported in part by the Research Council of Norway, 
Trieste University and MIUR.

\end{acknowledgments}


\vfill\eject
\end{document}